\documentstyle[emulateapj,onecolfloat,psfig]{article}

\def\spose#1{\hbox to 0pt{#1\hss}}
\def\lta{\mathrel{\spose{\lower 3pt\hbox{$\sim$}}
    \raise 2.0pt\hbox{$<$}}}
\def\gta{\mathrel{\spose{\lower 3pt\hbox{$\sim$}}
    \raise 2.0pt\hbox{$>$}}}
\def\arcsec{\hbox{$^{\prime\prime}$}}

\def\farcs{\hbox{$.\!\!^{\prime\prime}$}}

\begin{document}

\twocolumn[
\title{On the Origin of the Color-Magnitude Relation in the Virgo Cluster}
\author{Alexandre Vazdekis \altaffilmark{1,2},  
Harald Kuntschner \altaffilmark{1}, Roger. L. Davies \altaffilmark{1}}
\affil{
$^1$ Department of Physics, University of Durham, Durham DH1 3LE, U.K;
vazdekis@ll.iac.es, Harald.Kuntschner@durham.ac.uk, Roger.Davies@durham.ac.uk}
\affil{
$^2$ Instituto de Astrof\'{\i}sica de Canarias, 
E-38200 La Laguna, Tenerife, Spain}
\author{N. Arimoto \altaffilmark{3}, O. Nakamura \altaffilmark{3}}
\affil{$^{3}$ Institute of Astronomy, University of Tokyo, Osawa 2-21-1, Mitaka, 
Tokyo 181-0015, Japan; arimoto@ioa.s.u-tokyo.ac.jp, nakamura@ioa.s.u-tokyo.ac.jp}
\author{R. F. Peletier \altaffilmark{4}}
\affil{$^{4}$ School of Physics and Astronomy, University of Nottingham, Nottingham, 
NG7 2RD, U.K; Reynier.Peletier@nottingham.ac.uk}

\submitted{To appear in The Astrophysical Journal Letters, 2001 April 20 (551, number 2)}


\begin{abstract}
  We explore the origin of the color-magnitude relation (CMR) of early
  type galaxies in the Virgo cluster using spectra of very high S/N
  ratio for six elliptical galaxies selected along the CMR. The data
  are analysed using a new evolutionary stellar population synthesis
  model to generate galaxy spectra at the resolution given by their
  velocity dispersions. In particular we use a new age indicator that
  is virtually free of the effects of metallicity. We find that the
  luminosity weighted mean ages of Virgo ellipticals are greater than
  $\sim 8$~Gyr, and show no clear trend with galaxy luminosity.  We
  also find a positive correlation of metallicity with luminosity,
  color and velocity dispersion. We conclude that the CMR is driven
  primarily by a luminosity-metallicity correlation. However, not all
  elements increase equally with the total metallicity and we speculate
  that the CMR may be driven by both a total metallicity increase and
  by a systematic departure from solar abundance ratios of some
  elements along the CMR. A full understanding of the r\^ole played by
  the total metallicity, abundance ratios and age in generating the CMR
  requires the analysis of spectra of very high quality, such as those
  reported here, for a larger number of galaxies in Virgo and other
  clusters.

\end{abstract}

]

\section{Introduction}

To understand how galaxies form and evolve, we need to account for the
{\em color-magnitude relation} (CMR): luminous early-type galaxies in
clusters are observed to be redder than fainter ones (Visvanathan \&
Sandage 1977; Bower et al. 1992a). This tight relation, together with
the related Mg$_2$-$\sigma$ relation (Colless et al. 1999) links the
mass of a galaxy, through its luminosity, to its constituent stellar
populations.

The origin of the CMR is hotly debated: it could be caused either by a
variation of the mean stellar metallicity along the sequence (Arimoto
\& Yoshii 1987; Kodama \& Arimoto 1997), or by a combination of age
{\em and} metallicity variations (Gonz\'alez 1993; Ferreras, Charlot \&
Silk 1999). Larson (1974) suggested a dissipational collapse picture
for the formation of ellipticals with an early formation of the bulk of
the stars. In this scenario more luminous galaxies have greater binding
energies and can therefore achieve higher metallicities.
Terlevich~et~al. (1999) provided support for this view by showing that
the Coma cluster galaxies on the CMR are old, and that the galaxies
with bluer colors than expected for their luminosity are younger.
In more complex star formation histories, such as
implied by hierarchical merging (Cole et al. 1994; Kauffmann \& Charlot
1998) where galaxies have some recent star formation due to late merger
events, the CMR might be expected to arise from a combination of age
and metallicity variations. Indeed Gonz\'alez (1993), J{\o}rgensen
(1999) and Trager~et~al. (2000a) find strong evidence for a significant
intermediate age population in some elliptical
galaxies. Trager~et~al. (2000b) reported a large age range in their
full sample but note that the galaxies in clusters are generally
old. They also find a tight age-metallcity-velocity dispersion relation
such that at a fixed velocity dispersion, metal rich galaxies are young
and conjecture that this relation might give rise to a tight CMR.

Broad band colors and line strength indices of integrated stellar
populations are sensitive to both changes in age {\em and} metallicity and
this degeneracy has prevented an unambiguous analysis of the ages and
metallicities of ellipticals along the CMR (Worthey 1994). Even the
widely used age indicator in the Lick/IDS system, H$\beta$, shows a 
non-negligible dependence on metallicity (Worthey 1994) and can be 
affected by emission (Davies, Sadler \& Peletier 1993; Gonz\'alez 1993).
The origin of the CMR is not yet fully understood, in part, because of the
lack of appropriate tools of analysis.

In this Letter we study the stellar populations of six Virgo
ellipticals selected along the CMR for which we have very high S/N
spectra. We use a new stellar population synthesis model in combination
with new, more sensitive age indicators to derive age estimates that
are virtually independent of the metallicity.

\section{Stellar population Models}
\label{sec:models}

Here we use the evolutionary stellar population synthesis model of
Vazdekis (1999), which employs the extensive empirical
stellar spectral library of Jones (1999). The model predicts spectral
energy distributions (SEDs) in the optical wavelength range for simple
stellar populations (SSPs) at a resolution of
1.8~\AA\/ (FWHM). Previous models (e.g., Worthey 1994; Vazdekis
et al. 1996) used mostly the Lick polynomial fitting functions (Worthey
et al. 1994; Worthey \& Ottaviani 1997) to relate the strengths of
selected absorption features to stellar atmospheric parameters. The
fitting functions are based on the Lick/IDS stellar library
(FWHM$\sim$9~\AA, Worthey et al. 1994), thus limiting the application
of the models to strong features.  Furthermore the Lick stellar library
has not been flux calibrated, so offsets had to be applied to data
obtained with different spectrographs (Worthey \& Ottaviani 1997).

The new model provides flux calibrated, high resolution spectra,
therefore allowing an analysis of galaxy spectra at the natural
resolution given by their internal velocity broadening and instrumental
resolution. Furthermore, as the model outputs are SEDs rather than
predicted index strengths, 
it is easy to define new indices and even to
confront the model predictions with the detailed structure of observed
absorption features. Vazdekis \& Arimoto (1999)
defined a new age indicator [H$\gamma+\frac{1}{2}$(Fe{\sc I}+Mg{\sc
  I})]$_{\sigma}$ (hereafter H$\gamma_{\sigma}$), which is virtually
free from metallicity dependence. 
To cover a large range in $\sigma$, Vazdekis \& Arimoto defined
three slightly different indices which give stable and sensitive age
predictions within the $\sigma$ ranges quoted in the sub-indices:
H$\gamma_{100<\sigma<175}$, H$\gamma_{150<\sigma<225}$ and
H$\gamma_{225<\sigma<300}$.

We emphasise that the model spectra we are fitting have a single
metallicity and age. 
We are measuring luminosity-weighted properties,
which means that the presence of a young (i.e., bright) stellar
component will disproportionally affect the mean age and therefore will
be detected more easily.


\begin{figure}
\centerline{\psfig{file=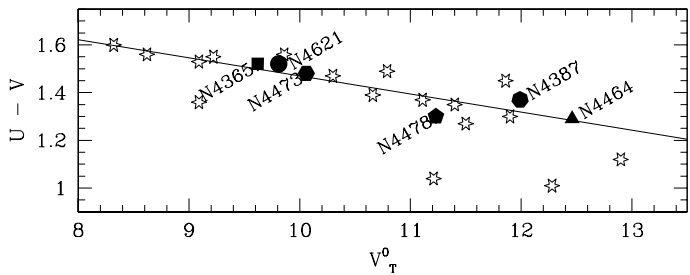,width=3.8in}}
\vskip-5.6cm
\caption{\label{fig:CMD}The CMR in the Virgo cluster. Data and a best-fit 
  regression (ellipticals only) from Bower, Lucey \& Ellis (1992a,b). For 
  NGC~4464 we have taken V$_T$ from RC3.  For NGC~4478 we used Michard (1982)
  $U-V$ color (60\arcsec aperture) applying an offset of $-0.047$,
  obtained by a comparison of his non dereddened $U-V$ colors with
  those of Bower, Lucey \& Ellis (1992b) which are dereddened by
  $-0.02$. Solid symbols represent our sample of galaxies, the shapes
  are used to identify the six galaxies in the following figures.}
\end{figure}

\section{Data}

We obtained major axis spectra of very high S/N ratios for six Virgo
ellipticals NGC~4365, NGC~4387, NGC~4464, NGC~4473, NGC~4478, and
NGC~4621, selected along the CMR of Bower, Lucey \& Ellis (1992a,b)
(see Fig.~\ref{fig:CMD}, and also Table~1). The observations were
performed at the WHT (4.2m) at the Observatorio del Roque de Los
Muchachos, La Palma, in April 21-22, 1999. We used the ISIS double-beam
spectrograph with the EEV12 CCD and R600B grating in the blue channel,
which provided a dispersion of $0.44$~\AA/pix and a spectral range of
$\lambda\lambda\sim$4000-5500~\AA. The slit width was set to 1\farcs6
giving a spectral resolution of 2.4~\AA\ (FWHM).  The seeing was
2-3\arcsec\ in the first night and $\sim$1\farcs5 in the second. The
CCD was binned $\times$3 in the spatial direction giving 0\farcs57 per
bin. Multiple exposures of 35~$\min$ length were obtained for each
galaxy.

Data reduction was done with standard IRAF packages.  
We made use of tungsten flatfields obtained for each galaxy pointing. 
The continuum shape was corrected to a relative
flux scale with spectrophotometric standards.

We summed up the spectra within $r_{e}$/10, where effective radii were
taken from Burstein~et~al. (1987).
Before summation the central galaxy spectra were carefully 
aligned with the rows of the array, and each row in spatial direction 
was corrected for the velocity shift due to rotation. 
Variations of $\sigma$ within $r_{e}$/10 were small ($<25$ km~s$^{-1}$) 
for all galaxies and therefore no correction was applied. 
Table~1 lists the S/N ratio per~\AA\/ achieved in the summed spectra 
around the H$\gamma$ feature.

\section{The ages and metallicities of Virgo ellipticals}

In this section (i) we compare the sensitivity and precision of
different age indicators, (ii) demonstrate that not all metals are
enhanced at the same rate in more luminous (higher dispersion) galaxies
and (iii) show that for the six Virgo ellipticals studied here
metallicity increases with luminosity, velocity dispersion and color
but no trend with age is apparent.

Our primary goal is to measure the ages of the central regions of the
galaxy sample along the CMR. We use different age indicators:
H$\gamma_{\sigma}$, H$\delta_F$ and H$\beta$ (as defined by 
Vazdekis \& Arimoto (1999),
Worthey \& Ottaviani 1997, and Worthey et al. 1994, respectively). 
Instead of applying $\sigma$
corrections to the line strength measurements we modified the model
predictions. We divided our sample into three groups containing galaxies
with similar velocity dispersions and smoothed the SEDs of 
Vazdekis (1999)
accordingly. We chose (a) $\sigma_{tot}=145$, (b) $\sigma_{tot}=190$
and (c) $\sigma_{tot}=260$~kms$^{-1}$ (where
$\sigma_{tot}^2=\sigma_{galaxy}^2+\sigma_{instr.}^2$). To minimize the
effect of small velocity dispersion differences the spectrum of
NGC~4387 was convolved with a Gaussian of $\sigma=85$~kms$^{-1}$ and
that of NGC~4621 with $\sigma=105$~kms$^{-1}$.

\begin{figure}[t]
\centerline{\psfig{file=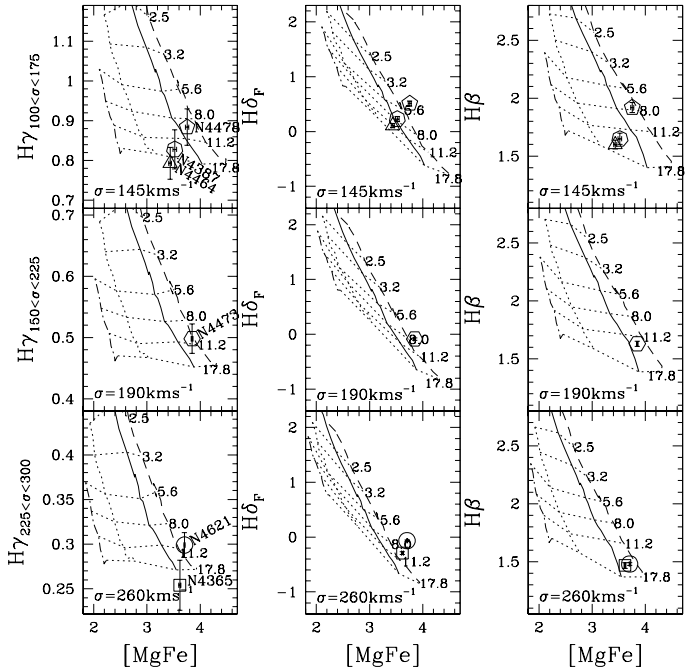,width=3.8in}}
\vskip-0.3cm
\caption{Plots of H$\gamma_{\sigma}$, H$\delta_F$ and H$\beta$ (from
  left to right) versus the metallicity index $[$MgFe$]$ (defined in
  Gonz\'alez 1993).  The central velocity dispersion increases in three
  groups from top to bottom, representing the CMR. Overplotted are the 
  models by Vazdekis (1999). Lines of
  constant [Fe/H]~$= -0.7$, $-0.4$, $0.0$ and $+0.2$ are shown by thick
  dot-dashed, thick dotted, thick solid and thick dashed lines,
  respectively. Thin dotted lines represent models of constant ages
  which are quoted in Gyr.}
\label{fig:age}
\end{figure}

Fig.~\ref{fig:age} shows age/metallicity diagnostic diagrams for the
age indicators as a function of the mean metallicity indicator [MgFe]
(Gonz\'alez 1993). Velocity dispersion increases in each column from
top to bottom. Note that for a given age and metallicity the strength
of the [MgFe] index is smaller for larger $\sigma$. The left panels of
Fig.~\ref{fig:age} show how well 
H$\gamma_{\sigma}$ is able to break
the age-metallicity degeneracy, much better than, e.g., H$\delta_F$.
The lines of constant age are essentially horizontal, which means that
a given measurement of H$\gamma_{\sigma}$ corresponds to a unique
age. H$\beta$ is also a good age indicator, the
new models predict that H$\beta$ is more sensitive
to age, and less sensitive to metallicity than models using the
Lick fitting functions.

Overall we find good agreement between the ages inferred from
H$\gamma_{\sigma}$ and H$\beta$. All our galaxies, 
except for
NGC~4478, are older than $\sim 11$~Gyr. NGC~4478 is 8-9~Gyr old
and has a $U-V$ color 0.06~mag bluer than other Virgo ellipticals of
equivalent luminosities (see Fig.~\ref{fig:CMD}). 

The middle panels of Fig.~\ref{fig:age} show that the age
estimates based on H$\delta_F$ remain strongly coupled with metallicity
estimates. (NB. the H$\gamma_F$ index is similar
and is not plotted.) The age estimates based on
H$\delta_F$ are younger than those derived from
H$\gamma_{\sigma}$ or H$\beta$ but we are unable to distinguish between
genuinely younger ages and the possible systematic effects of non-solar
abundance ratios. For example, using H$\delta_F$ or H$\gamma_F$ as the
age indicator one derives younger ages when Mg\,$b$ and/or Mg$_2$ are
used as the metal indicator than when the Fe lines alone are used
(Worthey 1998; Kuntschner 2000).

\begin{figure}[t]
\centerline{\psfig{file=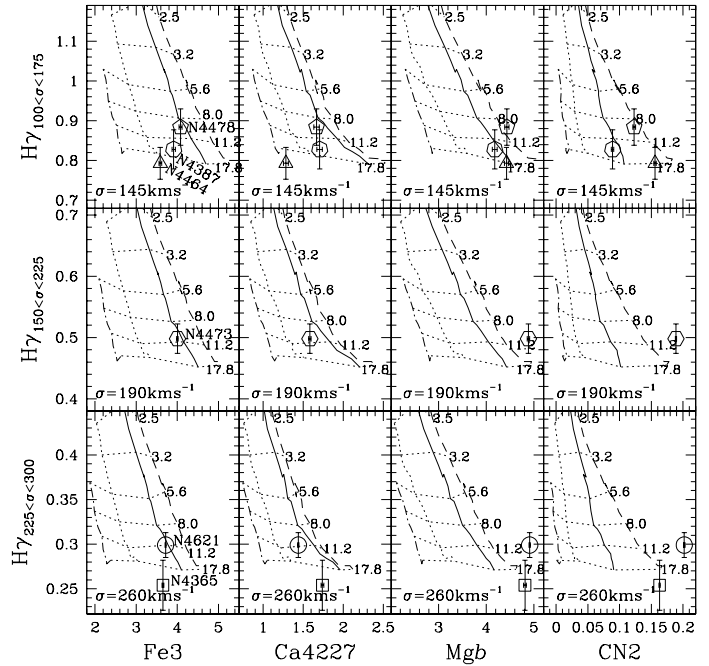,width=3.8in}}
\vskip-0.3cm
\caption{Plots of H$\gamma_{\sigma}$ {\em vs}\/ Fe3, Mg\,$b$, Ca4227
  and CN2. Symbols and line styles are the same as in Fig.~\ref{fig:age}.}
\label{fig:abund}
\end{figure}

Fig.~\ref{fig:abund} plots the H$\gamma_{\sigma}$ indices against the
strength of four metal absorption lines and illustrates the effects of
non-solar abundance ratios in determining metallicity. The left-hand
columns in Fig.~\ref{fig:abund} show only a very modest increase in the
metallicity derived from Fe3\footnote{Fe3~=~$\frac{1}{3}$(Fe4383 +
Fe5270 + Fe5335), Kuntschner (2000).} or Ca4227 as velocity dispersion
increases from top to bottom. We note that the G-band also follows
Fe3. 
In contrast the right-hand columns in Fig.~\ref{fig:abund}
show that the metallicity derived from Mg\,$b$ or CN2 increases
significantly towards high velocity dispersion galaxies, extending
beyond the model grids in some cases.  
Although Ca, like Mg, is an
$\alpha$-element, we find that Ca4227 does not track Mg\,$b$\/ 
confirming the results reported by Vazdekis et al. (1997), Worthey (1998) 
and Trager et al. (2000a). 
We conclude that there are at least two
families of metal lines which give rise to very different metallicity
estimates and exhibit different trends along the CMR. Despite these
complexities in estimating ``metallicity'', we emphasize that by using
H$\gamma_{\sigma}$ these uncertainies do not influence our derived ages.

\begin{figure}[t]
\centerline{\psfig{file=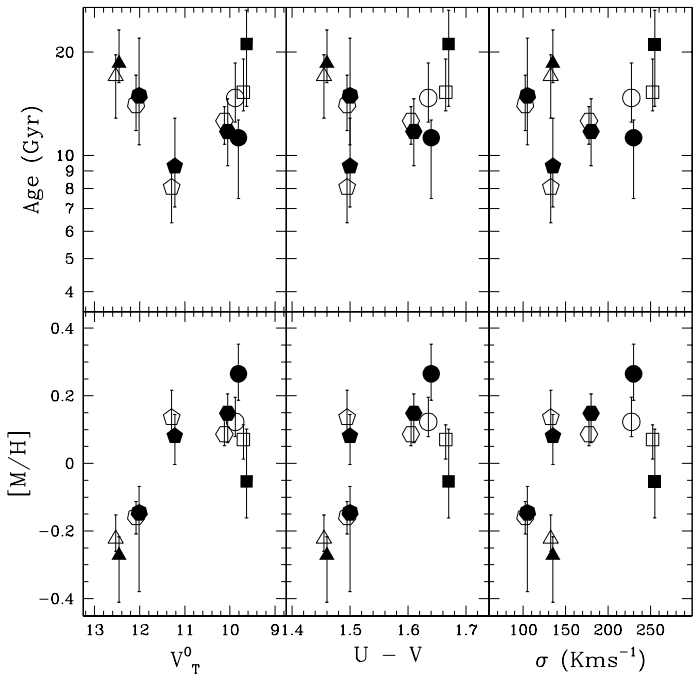,width=3.8in}}
\vskip-0.3cm
\caption{Mean galaxy ages and metallicities as estimated from
H$\gamma_{\sigma}$ (filled symbols) and H$\beta$ (open symbols)
(derived from H$\beta$ {\em vs} [MgFe] diagrams) {\em vs} V$^{0}_{T}$,
$U-V$ and velocity dispersion. The error bars for the
H$\gamma_{\sigma}$ age and metallicity estimates are mainly due to
noise, while those for H$\beta$ are dominated by the larger
age/metallicity degeneracy affecting this index.}
\label{fig:agemet}
\end{figure}

Fig.~\ref{fig:agemet} shows our estimates for luminosity weighted
mean age and metallicity (derived from H$\gamma_{\sigma}$ and H$\beta$
vs.  [MgFe] in Fig.~\ref{fig:age}) {\em versus} galaxy luminosity,
color, and velocity dispersion.  Except for NGC~4478, all ellipticals
are old with ages ranging from $\sim$11~Gyr to $\sim$20~Gyr. There is
no trend in age along the CMR but there is a clear, positive
correlation of metallicity with V$^{0}_{T}$, $U-V$ and $\sigma$. If we
correct the metallicity estimates to the value they would have if all
the galaxies were 14~Gyr old the scatter in Fig.~\ref{fig:agemet} is reduced.
Furthermore our [MgFe] 
index yields very similar metallicities to
those derived by correcting the Fe indices for non-solar abundance ratios,
following Trager et al. (2000a). We conclude that for these
six galaxies in Virgo the metal content increases along the CMR.

\section{Conclusions}

We have performed a detailed spectral analysis of six elliptical
galaxies in the Virgo cluster. We have combined very high S/N spectra
with a new evolutionary stellar population synthesis model. This model
enables us to study galaxy spectra at the resolution given by their
velocity dispersions and determine ages that are virtually free of any
dependence on metallicity. We find that elliptical galaxies in Virgo
are generally old.

We find that the metallicity as measured by [MgFe] correlates with
luminosity, $U-V$ and velocity dispersion. We conclude that the CMR in
Virgo is driven primarily by an increase in metallicity with
luminosity. We find that not all elements increase equally with the
total metallicity estimates. For example the increase in metallicity as
measured from the Fe or Ca indices is very modest along the CMR,
whereas the Mg or CN indices indicate a more rapid increase with galaxy
luminosity. This is consistent with the findings of Worthey, Faber, \&
Gonz\'{a}lez (1992), Kuntschner (2000) and Trager et al. (2000b) who
find an increase of the [Mg/Fe] ratio with increasing central velocity
dispersion. We speculate that the CMR in Virgo may be driven by both a
total metallicity increase and by a systematic departure from solar
abundance ratios of some elements along the CMR.

The luminosity weighted mean ages range from $\sim 8-20 $~Gyr but there
is no clear trend between age and V$^{0}_{T}$, $U-V$ or $\sigma$. 
Trager et al. (2000b) note that at a given velocity
dispersion galaxies with younger ages have higher metallicity. Our
Virgo galaxies fall on the Trager dispersion-age-metallicity relation
but span less than half the range of ages and metallcities that are
found in their sample. Trager et al. suggest that the distribution of
ages and metallicities amongst ellipticals may depend on environment
and indeed our Virgo galaxies exhibit a much smaller range than that
found in their mixed environment sample.  We cannot determine whether
the small age range we find here is universal in cluster ellipticals
but we intend to address this question in future work.
 
\acknowledgments The authors thank J. Blakeslee, R. Bower, J. Gorgas, T. Kodama, J.
Lucey and the referee, J.J. Gonz\'alez, for valuable discussions and suggestions.
The authors thank the referee for very useful suggestions. A.V. and H.K. acknowledge the
support of the PPARC rolling grant 'Extragalactic Astronomy and Cosmology in Durham
1998-2002'.  This work was financially supported in part by a Grant-in-Aid for Scientific
Research (No.1164032) by the Japanese Ministry of Education, Culture, Sports and
Science.  RLD acknowledges the support of a Leverhulme Trust Research Fellowship.

\begin{deluxetable}{ccccccc}
  \tablecaption{Virgo galaxy sample. \label{tbl-1}} \tablewidth{0pt}
  \tablehead{ \colhead{NGC} &\colhead{Exposure ($minutes$)} &\colhead{$r_{e}/10$}
    &\colhead{S/N per \AA} &\colhead{$\sigma$($kms^{-1}$)}
    &\colhead{$V_T^{0}$}\tablenotemark{a}
    &\colhead{$(U-V)^{0}_{10}$}\tablenotemark{b} } \startdata
  4464& 105& 0\farcs6& $\sim$160& $\sim$135& 12.46\tablenotemark{c}& 1.46\tablenotemark{a}\\
  4387& 105& 2\farcs6& $\sim$130& $\sim$105& 11.99& 1.50\\
  4478& 105& 1\farcs7& $\sim$135& $\sim$135& 11.23& 1.50\\
  4473&  70& 2\farcs7& $\sim$280& $\sim$180& 10.06& 1.61\\
  4621& 140& 6\farcs0& $\sim$490& $\sim$230&  9.81& 1.64\\
  4365&  70& 6\farcs7& $\sim$250& $\sim$255&  9.62& 1.67\\
  \enddata \tablenotetext{a}{Bower, Lucey \& Ellis (1992b).}
  \tablenotetext{b}{Michard (1982). $(U-V)^{0}_{10}$ corrected to
    $r_{e}/10$ from $(U-V)_{e}$ using the mean gradient of Trager et
    al. (2000b). Galactic extinction corrected using Schlegel et al.
    (1998).}  \tablenotetext{c}{RC3.}
\end{deluxetable}

%

\end{document}